\documentclass[aps,prc,preprint,superscriptaddress]{revtex4}

\usepackage{graphicx}
\usepackage{dcolumn}
\usepackage{bm}

\def\ucla{\affiliation{University of California Los Angeles, Los Angeles, California 90095-1547, USA}}
\def\umainz{\affiliation{Institut f\"ur Kernphysik, University of Mainz, D-55099 Mainz, Germany}}
\def\uglasgow{\affiliation{SUPA, School of Physics and Astronomy, University of Glasgow, Glasgow G12 8QQ, UK}}
\def\kentsu{\affiliation{Kent State University, Kent, Ohio 44242-0001, USA}}
\def\ubonn{\affiliation{Helmholtz-Institut f\"ur Strahlen- und Kernphysik, University of Bonn, D-53115 Bonn, Germany}}
\def\pnpi{\affiliation{Petersburg Nuclear Physics Institute, Gatchina 188350, Russia}}
\def\ubasel{\affiliation{Institut f\"ur Physik, University of Basel, CH-4056 Basel, Switzerland}}
\def\pavia{\affiliation{INFN Sesione di Pavia, I-27100 Pavia, Italy}}
\def\uedinburgh{\affiliation{School of Physics, University of Edinburgh, Edinburgh EH9 3JZ, UK}}
\def\gwu{\affiliation{The George Washington University, Washington, D.C. 20052, USA}}
\def\lebedevpi{\affiliation{Lebedev Physical Institute, Moscow, 119991, Russia}}
\def\ugiessen{\affiliation{II Physikalisches Institut, University of Giessen, D-35392 Giessen, Germany}}
\def\uma{\affiliation{Mount Allison University, Sackville, NB E4L 1E6, Canada}}
\def\rbu{\affiliation{Rudjer Boskovic Institute, Zagreb 10002, Croatia}}
\def\inr{\affiliation{Institute for Nuclear Research, Moscow 117312, Russia}}
\def\cua{\affiliation{The Catholic University of America, Washington, D.C. 20064, USA}}

\begin{document}


\title{Measurement of $3\pi^0$ photoproduction on the proton from threshold to 1.4~GeV}
\author{A.~Starostin}\email[]{starost@ucla.edu}\ucla
\author{I.~M.~Suarez}\altaffiliation[]{present address: Department of Physics, Texas A\&M University, 
College Station, Texas 77843-4242, USA}\ucla
\author{B.~M.~K.~Nefkens}\ucla
\author{J.~Ahrens}\umainz
\author{J.~R.~M.~Annand}\uglasgow
\author{H.~J.~Arends}\umainz
\author{K.~Bantawa}\kentsu
\author{P.A.~Bartolome}\umainz
\author{R.~Beck}\ubonn
\author{V.~Bekrenev}\pnpi
\author{A.~Braghieri}\pavia
\author{D.~Branford}\uedinburgh
\author{W.~J.~Briscoe}\gwu
\author{J.~Brudvik}\ucla
\author{S.~Cherepnya}\lebedevpi
\author{B.T.~Demisse}\gwu
\author{M.~Dieterle}\ubasel
\author{E.~J.~Downie}\umainz\uglasgow
\author{L.~V.~Fil'kov}\lebedevpi
\author{D.~I.~Glazier}\uedinburgh
\author{R.~Gregor}\ugiessen
\author{E.~Heid}\umainz
\author{D.~Hornidge}\uma
\author{I.~Jaegle}\ubasel
\author{O.~Jahn}\umainz
\author{T.~C.~Jude}\uedinburgh
\author{V.~L.~Kashevarov}\lebedevpi
\author{I.~Keshelashvili}\ubasel
\author{R.~Kondratiev}\inr
\author{M.~Korolija}\rbu
\author{A.~Koulbardis}\pnpi
\author{S.~Kruglov}\pnpi
\author{B.~Krusche}\ubasel
\author{V.~Lisin}\inr
\author{K.~Livingston}\uglasgow
\author{I.~J.~D.~MacGregor}\uglasgow
\author{D.~M.~Manley}\kentsu
\author{M.~Martinez}\umainz
\author{J.~C.~McGeorge}\uglasgow
\author{E.~F.~McNicoll}\uglasgow
\author{D.~Mekterovic}\rbu
\author{V.~Metag}\ugiessen
\author{A.~Mushkarenkov}\pavia
\author{M.~Oberle}\ubasel
\author{M.~Ostrick}\umainz
\author{P.~Pedroni}\pavia
\author{A.~Polonski}\inr
\author{S.~Prakhov}\ucla
\author{J.~Robinson}\uglasgow
\author{G.~Rosner}\uglasgow
\author{T.~Rostomyan}\ubasel
\author{S.~Schumann}\umainz\ubonn
\author{M.~H.~Sikora}\uedinburgh
\author{D.~Sober}\cua
\author{I.~Supek}\rbu
\author{M.~Thiel}\ugiessen
\author{A.~Thomas}\umainz
\author{M.~Unverzagt}\umainz\ubonn
\author{D.~P.~Watts}\uedinburgh
\author{D.~Wertmueller}\ubasel
\author{L.~Witthauer}\ubasel
\homepage[]{http://wwwa2.kph.uni-mainz.de}
\collaboration{Crystal Ball Collaboration at MAMI}
\date{\today}
\begin{abstract}
The total cross section for $\gamma p \to 3 \pi^0 p$ has been measured for the first time from threshold 
to 1.4~GeV using the tagged photon beam of the Mainz Microtron. The equipment utilized the
Crystal Ball multiphoton spectrometer, the TAPS forward detector and a particle identification
detector. The excitation function for $\sigma_{\rm total}(\gamma p \to 3 \pi^0 p)$ has two broad enhancements 
at $\sqrt{s} \approx 1.5$~GeV and 1.7~GeV. We obtained 
$\sigma_{\rm total}(\gamma p \to 3 \pi^0 p)/\sigma_{\rm total}(\gamma p \to \eta p) = 0.014 \pm 0.001$ at 
$\sqrt{s} \approx 1.5$~GeV.
\end{abstract}

\pacs{13.60.Hb, 13.60.Rj, 14.20.Gk}
\maketitle

\section{Introduction}
The spontaneous breaking of QCD chiral symmetry leads to formation of hadrons with mass. Restoration of chiral 
symmetry under certain conditions was theoretically predicted and has been looked for experimentally in nuclear 
media at normal density and in high-energy nuclear collisions. 
An experimental observation of mass degenerated chiral partners could be one of the signs of such restoration.
Meanwhile, mass degenerated chiral multiplets are observed in the spectrum of excited baryons where 
pairs of $N^\ast$ and $\Delta^\ast$ with the same angular momentum but opposite parities form parity doublets with 
nearly the same masses. Glozman was the first to point out that the existence of the parity doublets can be 
attributed to the restoration of chiral symmetry in highly excited hadronic states~\cite{glozman1,glozman2}. The 
third resonance region consists of three pairs 
of $N^\ast$ and two pairs of $\Delta^\ast$ (assuming the existence of the one-star $\Delta(1750)\frac{1}{2}^+$), 
which can be qualified as parity doublets, therefore the mass region around $\sqrt{s} \sim 1700$~MeV is 
particularly important for an experimental
investigation of baryonic chiral multiplets. Decoupling of the $N^\ast$ sextet that consists of
$N(1650)\frac{1}{2}^-$, 
$N(1675)\frac{5}{2}^-$, 
$N(1680)\frac{5}{2}^+$, 
$N(1700)\frac{3}{2}^-$, 
$N(1710)\frac{1}{2}^+$ and 
$N(1720)\frac{3}{2}^+$,
and the four $\Delta^\ast$ states 
($\Delta(1600)\frac{3}{2}^+$, 
$\Delta(1620)\frac{1}{2}^-$,
$\Delta(1750)\frac{1}{2}^+$ and
$\Delta(1700)\frac{3}{2}^-$)
is however a challenging experimental task. These states are broad and overlapping. In fact, two of the six $N^\ast$, 
namely $N(1700)\frac{3}{2}^-$ and $N(1710)\frac{1}{2}^+$, are not seen in the recent SAID partial-wave 
analysis (PWA)~\cite{said06}.
As the coupling of $N^\ast$ and $\Delta^\ast$ to $\pi N$ decreases with the increase of the incident energy, new 
information on decay modes other than $N^\ast \to \pi N$ is needed to determine the properties of the states. We 
have addressed this issue previously, publishing new high precision 
$\gamma p \to \eta p$ data~\cite{etatcs_sergey}. In this paper we present the first measurement
of the $\gamma p \to 3 \pi^0 p$ reaction, which has been made at the Mainz Microtron facility (MAMI).

The expected production process can be described by the isobar model, for example,
in the second resonance region $\gamma p \to N_1^\ast \to N_2^\ast \pi_1^0$ followed by 
$N_2^\ast \to \Delta(1232) \pi_2^0$ where $\Delta(1232)$ decays to $\pi_3^0 p$. The alternative decay branch is 
$N_2^\ast \to f_0(600) p$ where $f_0(600)$ decays to $\pi_2^0 \pi_3^0$. 
Similar decay chains are expected in the third resonance region, where $N^\ast_1$ can be one of the sextet $N^\ast$ 
states or one of the four $\Delta^\ast$ states mentioned above, but additional decay modes are allowed for the 
$\Delta^\ast$ states. For example, $\gamma p \to \Delta^\ast  \to f_0(600) \Delta(1232)$ followed by 
$f_0(600) \to \pi^0_1 \pi^0_2$ and $\Delta(1232) \to \pi^0_3 p$.
Previously the total cross section for the related pion production process, $\pi^-p \to 3 \pi^0 n$,
has been used successfully to determine the branching ratio of 
$N(1535)\frac{1}{2}^- \to N(1440)\frac{1}{2}^+ \pi$~\cite{cbags_3pi0}.

\section{Experiment and data analysis}
The data were collected in 2009 using the upgraded MAMI-C electron microtron that provides a high
intensity beam of electrons with maximum energy of 1557~MeV. The MAMI-Glasgow tagging 
spectrometer~\cite{anthony,hall,mcgeorge} was used to produce a secondary beam of tagged photons
with maximum energy of $\approx 1450$~MeV and energy resolution about $\pm 2$~MeV. The 
Bremsstrahlung beam was
incident on a 10-cm-long liquid hydrogen target installed in the center of the Crystal Ball (CB). The spectrometer 
is a nearly $4\pi$-coverage full absorption spherical calorimeter made
of 672 NaI(Tl) triangular-pyramidal crystals arranged in the form of two hemispheres. The crystals are
16 radiation lengths thick and the detector provides a typical energy resolution for photons
of $\sigma/E = 1.7\%/[E({\rm GeV})]^{0.4}$. The typical time resolution is 6~ns (FWHM). The
detector has two $21^\circ$ openings, one on the upstream and the other on the 
downstream side, and
a spherical cavity in the center of the sphere allowing installation of the liquid hydrogen target
assembly and the particle identification detector (PID) in the middle of the spectrometer. The PID
detector surrounding the liquid hydrogen target is made of 24 strips of plastic scintillator 50~cm
long and 4~mm thick equipped with a photomultiplier at the upstream end. Although the Crystal Ball
is optimized for the detection of photons and electrons, it also has good efficiency for detecting
neutrons and protons. See Refs.~\cite{cb1,cb2,cb3} and references wherein for a detailed description of the detector. 

The geometrical acceptance of the setup was increased by covering the downstream opening in the CB 
by the TAPS BaF$_2$ forward wall. In its current configuration, TAPS~\cite{novotny,gabler}
is comprised of 360 individual barium fluoride (BaF$_2$) crystals arranged in the form of a hexagonal prism.
Each TAPS crystal is shaped as a hexagon with an inscribed diameter of 59~mm and length 
250~mm (12 {\it r.l.}). A time resolution of 160~ps was achieved for the TAPS 
detector~\cite{novotny}. One
crystal in the center of the TAPS detector was removed to allow the beam to exit. The two inner
layers of the TAPS BaF$_2$ crystals (24 crystals in total) were replaced with 72 faster PbWO$_4$
crystals. However, the PbWO$_4$ detectors were not used in this analysis. The TAPS assembly is located
147.5~cm downstream of the center of the CB and covers the range in polar angle approximately between
$1^\circ$ and 20$^\circ$. The combination of CB and TAPS detectors covers $\approx 97$\% of the solid
angle.

The $\gamma p \to 3 \pi^0 p$ events were reconstructed from the pool of six-cluster (six photons only)
and seven-cluster (six photons and the proton cluster) events. Both Crystal Ball and TAPS clusters
were used in the analysis. A CB cluster consists of the central crystal (the crystal with the maximum
deposited energy) and up to 22 surrounding crystals with energies above 1.1~MeV. The energy in the
central crystal was required to be above 15~MeV. A TAPS cluster was constructed from the central
crystal, which has a minimum energy of 20~MeV, plus up to 18 surrounding crystals with energies
3.5~MeV or higher. The timing information for each crystal was used to ensure that all the hits in a
cluster originate from the same particle. The timing coincidence window for the Crystal Ball was set
to 70~ns, and for TAPS it was 30~ns. The energy of a photon cluster, calculated as the sum of the
energies deposited in all crystals contributing to the cluster, was corrected for the leakage of the
electromagnetic shower outside of the cluster boundaries and for nonlinearity of the analog-to-digital
converters. The energy-dependent corrections were calculated from a Monte Carlo simulation and
verified using the experimental data. For the case of the proton cluster, only the angular information
was used.

The detected events were subjected to a fit with kinematical constraints as described in earlier
Crystal Ball publications~\cite{cbags_eta, blobel}. All 15 combinations of three photon pairs from 
six-cluster events, and 105 combinations of three pairs plus one proton for seven-cluster events 
were tested against the $\gamma p \to 3 \pi^0 p \to 6 \gamma p$ hypothesis. The combination with the 
best $\chi^2$ satisfying the $\gamma p \to 3 \pi^0 p \to 6 \gamma p$ hypothesis at the 95\% 
confidence level (C.L.) was used for further analysis. Therefore the fit was exclusively used to 
determine the proton cluster and the three pairs of photon clusters from the $\pi^0$ decays. 
The efficiency of the method in respect to the combinatorial background is found to be better 
than 95\%. It was evaluated using Monte Carlo events. An event with $n$ hits in 
the beam tagger was treated as $n$ independent events with beam energies $E_n$.

\begin{figure}
\includegraphics[width=0.45\textwidth]{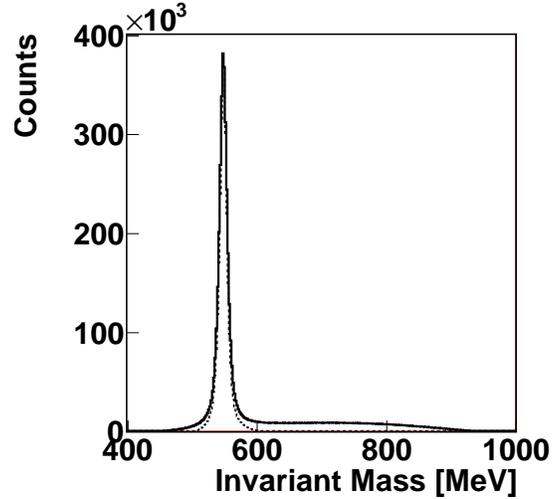}
\caption{Invariant mass spectrum of $3 \pi^0$ measured in the reaction 
$\gamma p \to 3 \pi^0 p \to 6 \gamma p$ and integrated over the photon beam energy range from
threshold to 1.43~GeV. The dominant peak is due to 
$\gamma p \to \eta p \to 3 \pi^0 p$ and the smooth background under the peak comes from the
$\gamma p \to 3 \pi^0 p$ direct production. The width of the $\eta$ peak has $\sigma = 6.9$~MeV.
The peak is compared to the results of the Monte Carlo simulation shown by the dashed line. 
\label{threepi0_invmass1}}
\end{figure}

\begin{figure}
\includegraphics[width=0.45\textwidth]{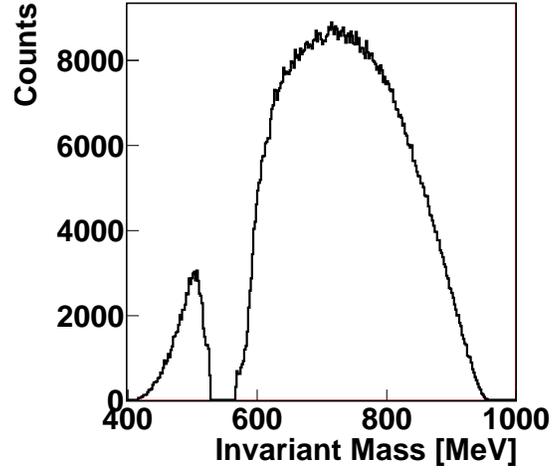}
\caption{Invariant mass spectrum of $3 \pi^0$ from $\gamma p \to 3 \pi^0 p \to 6 \gamma p$ 
after subtraction of the $\gamma p \to \eta p \to 3 \pi^0 p$ background and other backgrounds; 
see text for details.
\label{threepi0_invmass2}}
\end{figure}

Above $E_\gamma = 709$~MeV 
the dominant background for $\gamma p \to 3 \pi^0 p$ is the $\eta$ photoproduction reaction 
followed by the $\eta \to 3 \pi^0$ decay. In order to handle this background the events were 
also tested against the $\gamma p \to \eta p \to 3 \pi^0 p \to 6 \gamma p$ hypothesis. An event 
was rejected if it satisfied the $\gamma p \to \eta p \to 3 \pi^0 p \to 6 \gamma p$ hypothesis at 
the 99\% C.L. Figure \ref{threepi0_invmass1} shows the invariant mass of the reconstructed 
$3 \pi^0$ from the reaction $\gamma p \to 3 \pi^0 p$ integrated over all photon beam energies. 
The dominant peak in the distribution from 
$\gamma p \to \eta p \to 3 \pi^0 p \to 6 \gamma p$ is compared to the Monte Carlo simulation of 
$\gamma p \to \eta p \to 3 \pi^0 p \to 6 \gamma p$. The flat distribution under the peak is due 
to the direct $\gamma p \to 3 \pi^0 p$ reaction, which is the subject of this paper. The main part 
of the $\gamma p \to \eta p \to 3 \pi^0 p$ background was removed using the procedure described 
above. The residual $\eta \to 3 \pi^0$ background was determined by analyzing the 
$\gamma p \to \eta p \to 3 \pi^0 p$ Monte Carlo events as direct $\gamma p \to 3 \pi^0 p$.
This part of the background was normalized to the ratio of the reconstructed 
$\gamma p \to \eta p \to 3 \pi^0 p$ events in the simulation to the number of 
$\gamma p \to \eta p \to 3 \pi^0 p$ detected in the experiment and also subtracted. 
Additional corrections included subtraction of the good events associated with the accidental 
hits in the beam tagger ($\sim 13$\%, see Ref.~\cite{mdm_sven} for a detailed description of the
procedure) and subtraction of the empty target contribution, which ranges from about 50\% close 
to the threshold to about 3\% at higher beam energies. The invariant mass of the directly
photoproduced  $3 \pi^0$ after removal of the $\eta \to 3 \pi^0$ background, the random beam
background and the empty target contribution is shown in Fig.~\ref{threepi0_invmass2}.

\begin{figure}
\includegraphics[width=0.45\textwidth]{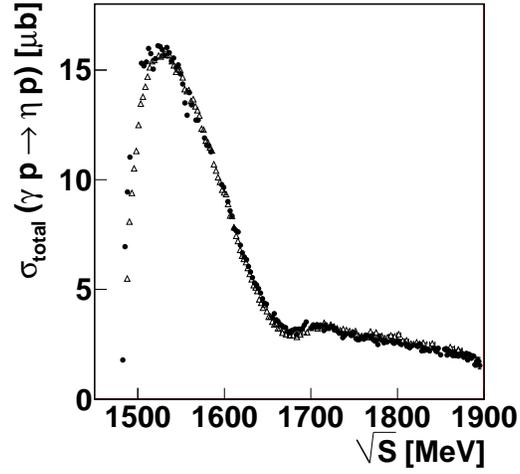}
\caption{The $\gamma p \to \eta p$ total cross section determined in this work 
(black dots) compared to the results from Ref.~\cite{etatcs_sergey} (open triangles).
\label{eta3pi0_tcs}}
\end{figure}

The number of photons in the beam was determined from the number of counts in the photon tagger
corrected to the tagger efficiency determined in a series of separate measurements, and to the live 
time of the data acquisition system; see Ref.~\cite{etatcs_sergey} for details. In order to verify 
the absolute normalization of the $\gamma p \to 3 \pi^0 p$ total cross section we have calculated 
the well-known $\gamma p \to \eta p$ total cross section. A comparison of our 
$\sigma_{\rm total}(\gamma p \to \eta p)$ 
to the results obtained in an analysis dedicated to $\gamma p \to \eta p$~\cite{etatcs_sergey} 
in shown in Fig.~\ref{eta3pi0_tcs}. 
Our results are within $\pm 5$\% agreement with the previously obtained $\gamma p \to \eta p$ 
cross sections.

\section{Results}
\begin{figure}
\includegraphics[width=0.45\textwidth]{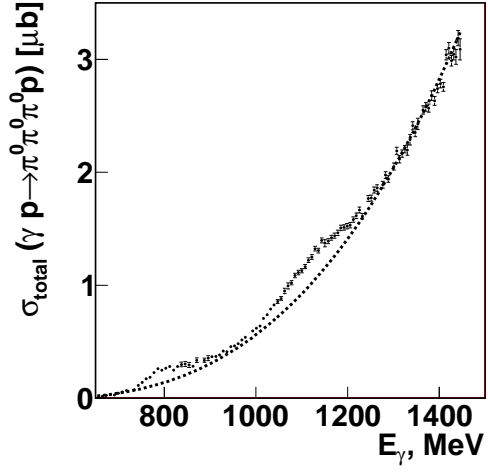}
\caption{The $\gamma p \to 3 \pi^0 p$ total cross section obtained in our experiment. The dotted line 
indicates a polynomial fit to the experimental data; see text for details.
\label{threepi0_tcs}}
\end{figure}

The measured $\gamma p \to 3 \pi^0 p$ total cross section is shown in 
Fig.~\ref{threepi0_tcs} and listed in Table~\ref{threepi0_tabl}. Only statistical uncertainties 
are shown. The systematical uncertainty is estimated to be 15\%. The three main sources of 
the systematical uncertainty are: (i) the uncertainty in the acceptance calculation associated 
with complex dynamics of the $\gamma p \to 3 \pi^0 p$ reaction; (ii) the remaining 
$\gamma p \to \eta p \to 3 \pi^0 p$ background for $E_{\gamma} > 710$~MeV;  (iii) the background 
from $\gamma p \to K_s^0 \Sigma^+$ followed by the decays $K_s^0 \to 2 \pi^0$ 
and $\Sigma^+ \to \pi^0 p$. The $\gamma p \to K_s^0 \Sigma^+$ background contributes at 
$E_{\gamma} > 1.05$~GeV and has not been subtracted. 

\begin{figure}
\includegraphics[width=0.9\textwidth]{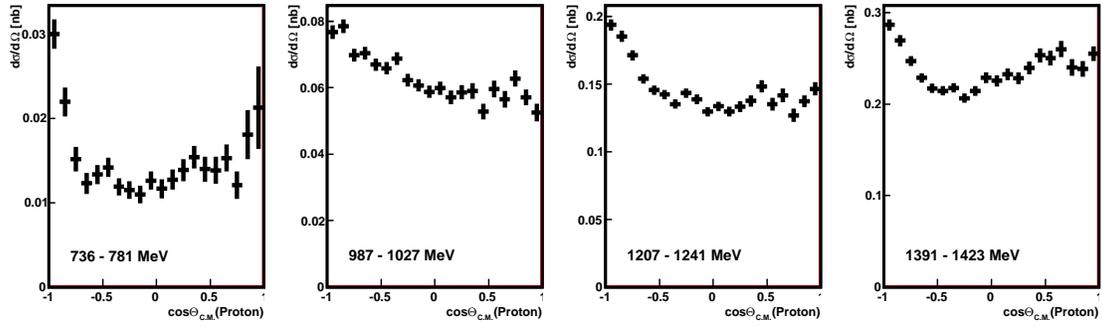}
\caption{Differential cross sections $d\sigma/d\Omega$ for the $\Theta_{p}$ in the center-of-mass 
frame for four beam energy intervals.
\label{threepi0_theta}}
\end{figure}

The uncertainty due to the acceptance calculation is estimated to be 5\%. The number was obtained 
from a comparison of $\sigma_{\rm total}(\gamma p \to 3 \pi^0 p)$ calculated from the integrated 
number of events to the total cross section obtained by integration of the differential distributions 
shown in Fig.~\ref{threepi0_theta}.
The $\gamma p \to \eta p \to 3 \pi^0 p$ and $\gamma p \to K_s^0 \Sigma^+$ backgrounds depend strongly 
on the magnitude of the corresponding total cross sections and therefore vary with the beam energy. 
However, by our estimate the sum of the two backgrounds is roughly invariant of the beam energy and 
does not exceed 10\% of the number of good $\gamma p \to 3 \pi^0 p$ events. 

The total cross section exhibits a smooth behavior from threshold ($\approx 492$~MeV) to 1.434~GeV 
where it reaches 2.9~$\mu$b. There is a local maximum of $0.25 \pm 0.01$~$\mu$b at about 
$E_\gamma = 780$~MeV. This can be compared to the total 
cross section for $\eta$ photoproduction at the peak of the $N(1535)\frac{1}{2}^-$ resonance: 
$\sigma_{\rm total}(\gamma p \to \eta p) = 16.0 \pm 0.1$~$\mu$b at 
$E_\gamma = 780$~MeV~\cite{etatcs_sergey}. The ratio of these two values is
\begin{equation}
R_{\gamma} = \frac{\sigma_{\rm total}(\gamma p \to 3 \pi^0 p)}{\sigma_{\rm total}(\gamma p \to \eta p)} = 0.016 \pm 0.001
\end{equation}
The corresponding ratio for the pion induced reactions measured in the earlier Crystal Ball 
experiments~\cite{cbags_3pi0,cbags_eta} is
\begin{equation}
R_{\pi} = \frac{\sigma_{\rm total}(\pi^- p \to 3 \pi^0 n)}{\sigma_{\rm total}(\pi^- p \to \eta n)} = \frac{0.023 \pm 0.004}{2.63 \pm 0.02} = 0.009 \pm 0.002.
\end{equation}
 
The two ratios have very similar values. This can be seen as an indirect confirmation of the
earlier estimate ${\cal B}(S_{11}(1535) \to P_{11}(1440) \pi) \approx 0.08$~\cite{cbags_3pi0}. 
The slight difference between the two ratios can possibly be attributed to the different interference 
between $N(1520)\frac{3}{2}^-$ and $N(1535)\frac{1}{2}^-$ and to the difference in 
the non-resonant background in photo- and pion production.

The isospin--conjugated $3 \pi$ photoproduction processes were measured previously at 
higher beam energies in several bubble chamber experiments~\cite{ahhmc_data,slac_data}. In particular,
the $\gamma p \to \pi^+ \pi^- \pi^0 p$ total cross section grows rapidly to about 20~$\mu$b
from threshold to $E_\gamma = 2$~GeV and then declines gradually to about 10~$\mu$b at $E_\gamma = 6$~GeV 
forming a broad peak with a maximum at around 2~GeV.
Assuming similar behavior for the $\gamma p \to 3 \pi^0 p$ cross section we can speculate that
the rapid growth of the $3 \pi^0$ cross section from threshold to 1.4~GeV is determined by the 
contributions from $N^\ast$ and $\Delta^\ast$ with masses 1.9--2.2~GeV/$c^2$. 

\begin{figure}
\includegraphics[width=0.45\textwidth]{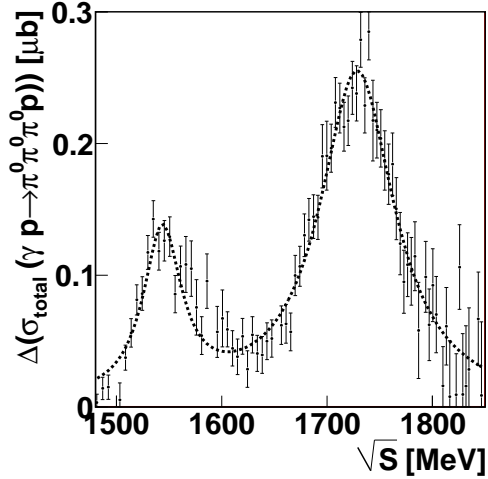}
\caption{The difference between the $\gamma p \to 3 \pi^0 p$ total cross section and the result of 
the polynomial fit shown on the previous figure. The dotted line shows the fit of the difference with 
two Breit-Wigner distributions. The peak positions from the fit are 1.53~GeV and 1.74~GeV.
\label{threepi0_tcs_diff}}
\end{figure}

\begin{figure}
\includegraphics[width=0.9\textwidth]{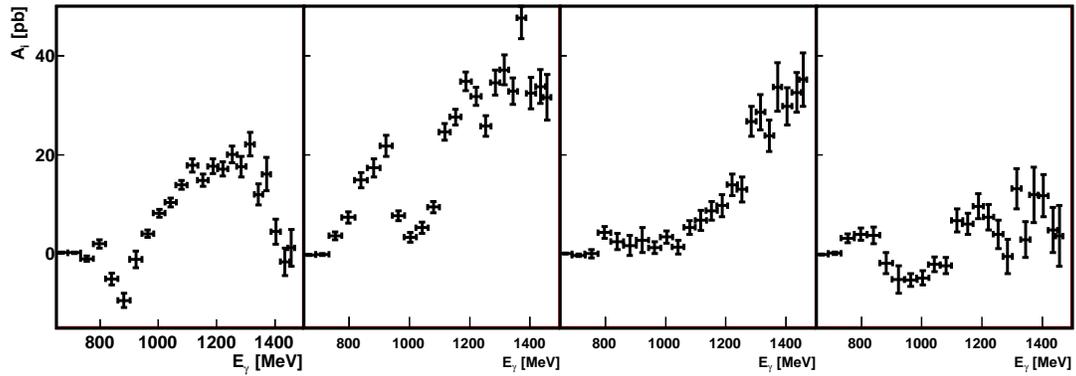}
\caption{Legendre coefficients $A_1 - A_4$ (from left to right) from the fit to $d\sigma/d\Omega$ 
for the $\Theta_{p}$ differential cross section.
\label{threepi0_legendre}}
\end{figure}

In order to study in more detail the enhancements at $\sqrt{s} \approx 1.5$~GeV and 1.7~GeV
we have fitted the cross section with a third--degree polynomial excluding the enhanced areas from the fit. 
The result of the fit is shown in Fig.~\ref{threepi0_tcs} by the dashed line. 
The part of the distribution above the fit curve is shown in Fig.~\ref{threepi0_tcs_diff}. 
A Breit-Wigner fit of the two peaks in Fig.~\ref{threepi0_tcs_diff} gives the positions of the peaks 
$\sqrt{s} = 1.54$~GeV and 1.73~GeV and the widths $\Gamma = 43$~MeV and 83~MeV,
respectively. The enhancements can possibly be attributed to the interference between the excited states 
in the second and the third resonance regions, although the structure at $\sqrt{s} = 1.54$~GeV can be 
partially due to the remaining $\gamma p \to \eta p$ background. It is worth noting that the 
$\gamma p \to \pi^0 \pi^0 p$
total cross section exhibits two similar enhancements with masses 50-70~MeV lower than the ones observed in 
our experiment~\cite{cbarrel_2pi0}.
Figure~\ref{threepi0_theta} shows the differential cross section for the angular distribution of the recoil 
proton in $\gamma p \to 3\pi^0 p$ in the center-of-mass frame for four beam energy intervals. The distributions 
all show a quite strong angular dependence. We fitted the angular distributions
with expansions in terms of Legendre polynomials. Figure~\ref{threepi0_legendre} shows the dependence of the 
Legendre coefficients $A_1 -A_4$ on the beam energy. The distributions of all coefficients 
show substantial structure.

\begin{figure}
\includegraphics[width=0.9\textwidth]{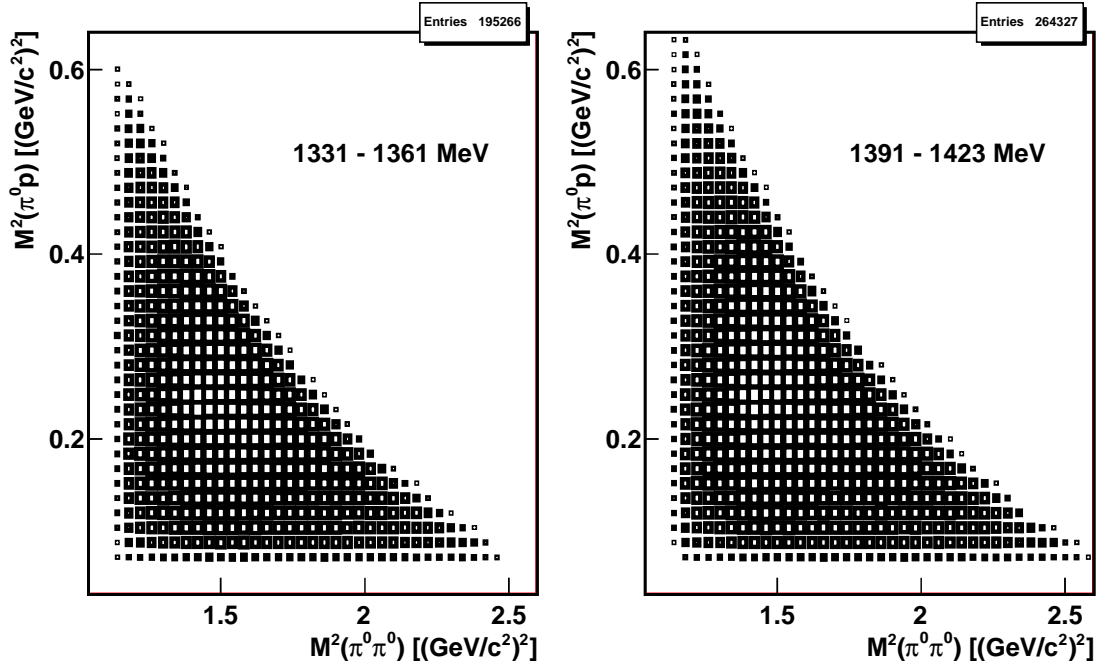}
\caption{Examples of the $\gamma p \to 3 \pi^0 p$ Dalitz plots. The $x$-axis shows $M^2(\pi^0 p)$ and the 
$y$-axis $M^2(\pi^0 \pi^0)$. Each events has three entries in the distribution representing different 
$\pi^0 p$ and $\pi^0 \pi^0$ combinations.
\label{threepi0_dp}}
\end{figure}

Further information on the dynamics of the $3 \pi^0$ photoproduction can be extracted from the Dalitz plot 
distributions. Examples of such distributions are shown in Fig.~\ref{threepi0_dp}. The axes of 
the Dalitz plot are $M^2(\pi^0 p)$ and $M^2(\pi^0 \pi^0)$. Each event has three entries on the
distribution representing three different combinations of $\pi^0 p$ and $\pi^0 \pi^0$.

\begin{figure}
\includegraphics[width=0.9\textwidth]{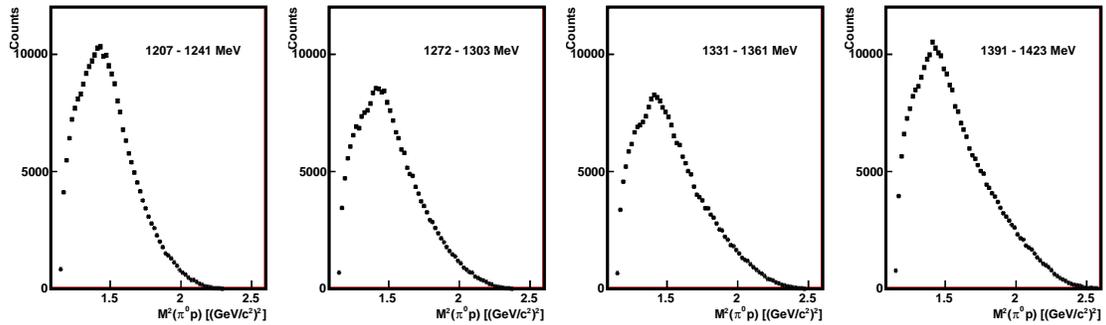}
\caption{Projection of the Dalitz plot onto the $M^2(\pi^0 p)$ axis for four beam energy intervals.
\label{threepi0_dpx}}
\end{figure}

\begin{figure}
\includegraphics[width=0.9\textwidth]{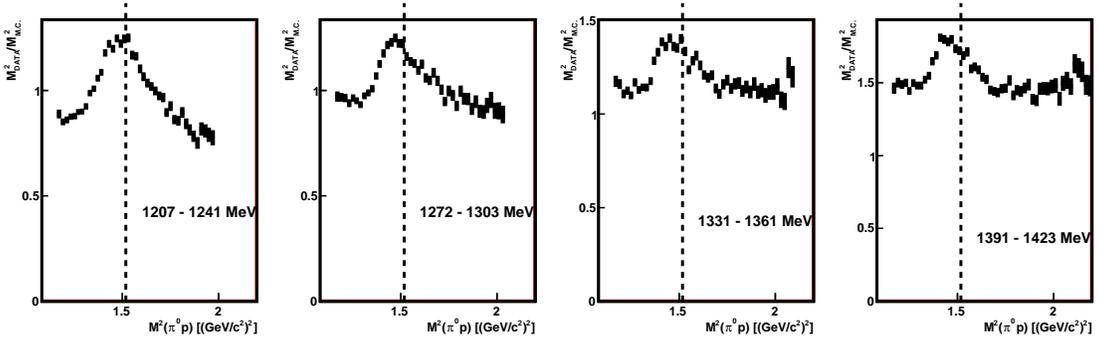}
\caption{Ratio of the $M^2(\pi^0 p)$ projections of the Dalitz plot obtained from the experimental 
data to the simulated distribution generated according to phase-space. The dashed line on the plots 
indicates the $\Delta (1232)$.
\label{threepi0_dp_rtx}}
\end{figure}

The Dalitz plot reflects the complex dynamics of the reaction. We expect that the $\pi^0 p$ interactions
lead to $\Delta(1232)$ formation, which shows up in Fig.~\ref{threepi0_dpx} as a strong peak near 1.5~GeV$^2/c^4$.
The $\Delta(1232)$ contribution is also seen in the ratio of the 
$M^2(\pi^0 p)$ projection of the Dalitz plot obtained from the data to the results of the Monte Carlo 
simulation generated according to phase-space
distribution. The ratios are shown in Fig.~\ref{threepi0_dp_rtx} for four beam intervals. The contribution from the 
$\Delta(1232)$ is also seen at around 1.5~GeV$^2/c^4$. The mass of the $\Delta$ peak on the projection is slightly off.
This is a combinatorial effect due to the fact that all three combinations of $M^2(\pi^0 p)$ are plotted.

\begin{figure}
\includegraphics[width=0.9\textwidth]{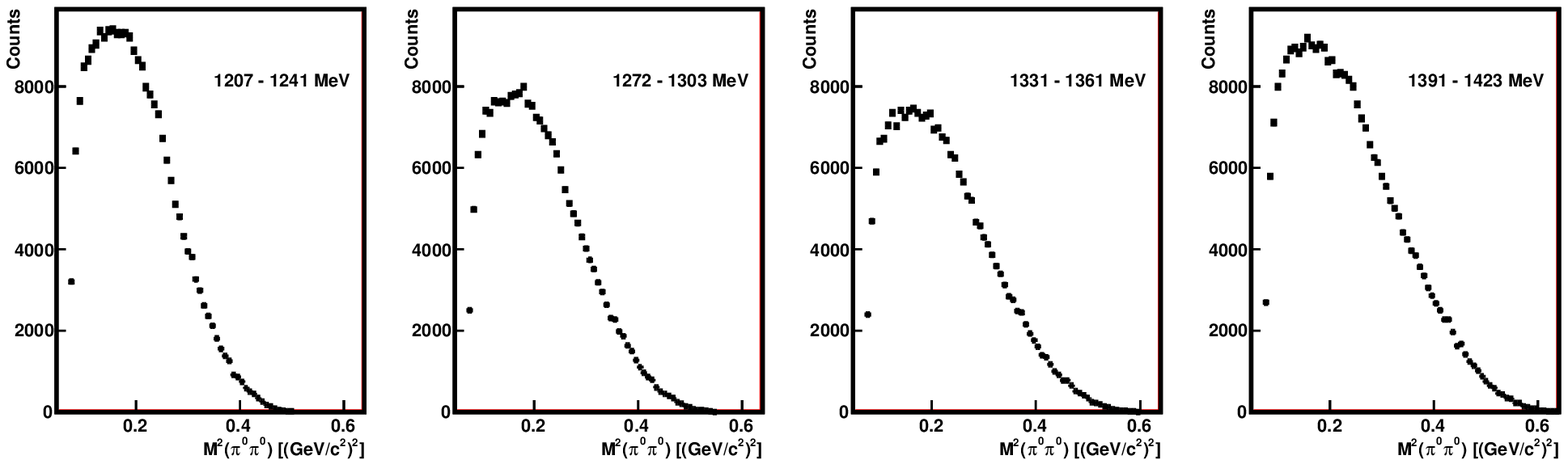}
\caption{Projection of the $\gamma p \to 3 \pi^0 p$ Dalitz plot unto the $M^2(\pi^0 \pi^0)$ axis for 
four beam energy intervals.
\label{threepi0_dp_pry}}
\end{figure}

The $\pi^0 \pi^0$ interaction leads to the intermediate $f_0(600)$ resonance. Its mass is not well known.
The projections of the Dalitz plot to $M^2(\pi^0 \pi^0)$ axis do not show any substantial structures in 
comparison to the phase-space distribution, see Fig.~\ref{threepi0_dp_pry}.
These features have been seen in the reaction $\pi^- p \to \pi^0 \pi^0 n$~\cite{bnl_2pi0} as well.
No substantial contribution from $f_0(600) \to 2 \pi^0$ was observed.

To summarize our results, the total cross section for $\gamma p \to 3 \pi^0 p$ has been measured 
for the first time from threshold to 1.4~GeV. 
The excitation function for $\sigma_{\rm total}(\gamma p \to 3 \pi^0 p)$ has two broad enhancements at 
$\sqrt{s} \approx 1.5$~GeV and 1.7~GeV. The ratio 
$\sigma_{\rm total}(\gamma p \to 3 \pi^0 p)/\sigma_{\rm total}(\gamma p \to \eta p) = 0.014 \pm 0.001$ was 
obtained at $\sqrt{s} \approx 1.5$~GeV confirming previous results for 
${\cal B}(S_{11}(1535) \to P_{11}(1440) \pi)$.

\section{Acknowledgments}
The success of the experiment on the $3 \pi^0$ photoproduction has been made possible by the use 
of the Crystal Ball neutral meson spectrometer with its 97\% acceptance per photon. 
We thank SLAC for letting us use the Crystal Ball.
The authors very much appreciate the dedicated work of the MAMI accelerator group. We also 
thank the undergraduate students of Mount Allison University and the George Washington University 
for their assistance in data taking. This work was supported by DOE and NSF of the U.S., EPSRC 
and STFC of the U.K., NSERC of Canada, the Deutsche Forschungsgemeinschaft (SFB 443) of Germany, 
and DFG-RFBR (Grant No.09-02-91330) of Germany and Russia, SNF of Switzerland and the European
Community-Research Infrastructure Activity under the FP6 ``Structuring the European Research 
Area'' program (Hadron Physics, contract number RII3-CT-2004-506078). 
%
\bibliography{prc_3pi0}
\begin{table}
\caption{$\gamma p \to 3 \pi^0 p$ total cross section. Only statistical uncertainties are 
listed.
\label{threepi0_tabl}}
\begin{ruledtabular}
\begin{tabular}{cc||cc}
   $E_{\gamma}$(MeV/$c$)  & $\sigma_{\rm total}(\gamma p \to 3 \pi^0 p)$($\mu$b)  &
   $E_{\gamma}$(MeV/$c$)  & $\sigma_{\rm total}(\gamma p \to 3 \pi^0 p)$($\mu$b)  \\
\colrule
           520 $\pm$  9 &                     $<$0.02   &         1044 $\pm$  8 &       0.841   $\pm$   0.012 \\ 
           538 $\pm$  9 &                     $<$0.02   &         1059 $\pm$  8 &       0.905   $\pm$   0.014 \\ 
           555 $\pm$  9 &                     $<$0.02   &         1074 $\pm$  8 &       1.012   $\pm$   0.015 \\ 
           573 $\pm$  9 &       0.002   $\pm$   0.001   &         1089 $\pm$  7 &       1.099   $\pm$   0.013 \\ 
           591 $\pm$  9 &       0.002   $\pm$   0.001   &         1104 $\pm$  7 &       1.147   $\pm$   0.013 \\ 
           608 $\pm$  9 &       0.003   $\pm$   0.001   &         1118 $\pm$  7 &       1.236   $\pm$   0.015 \\ 
           626 $\pm$  9 &       0.005   $\pm$   0.001   &         1133 $\pm$  7 &       1.315   $\pm$   0.015 \\ 
           643 $\pm$  9 &       0.009   $\pm$   0.002   &         1147 $\pm$  7 &       1.389   $\pm$   0.018 \\ 
           661 $\pm$  9 &       0.014   $\pm$   0.002   &         1161 $\pm$  7 &       1.405   $\pm$   0.015 \\ 
           678 $\pm$  9 &       0.027   $\pm$   0.002   &         1175 $\pm$  7 &       1.452   $\pm$   0.016 \\ 
           696 $\pm$  9 &       0.035   $\pm$   0.003   &         1189 $\pm$  7 &       1.510   $\pm$   0.017 \\ 
           713 $\pm$  9 &       0.058   $\pm$   0.005   &         1202 $\pm$  7 &       1.527   $\pm$   0.017 \\ 
           730 $\pm$  9 &       0.074   $\pm$   0.013   &         1216 $\pm$  7 &       1.599   $\pm$   0.018 \\ 
           748 $\pm$  9 &       0.125   $\pm$   0.007   &         1229 $\pm$  7 &       1.638   $\pm$   0.018 \\ 
           765 $\pm$  9 &       0.178   $\pm$   0.009   &         1242 $\pm$  6 &       1.769   $\pm$   0.027 \\ 
           782 $\pm$  9 &       0.243   $\pm$   0.009   &         1255 $\pm$  6 &       1.808   $\pm$   0.019 \\ 
           799 $\pm$  9 &       0.256   $\pm$   0.011   &         1267 $\pm$  6 &       1.862   $\pm$   0.029 \\ 
           816 $\pm$  8 &       0.263   $\pm$   0.010   &         1280 $\pm$  6 &       1.933   $\pm$   0.021 \\ 
           833 $\pm$  8 &       0.286   $\pm$   0.012   &         1292 $\pm$  6 &       1.940   $\pm$   0.029 \\ 
           850 $\pm$  8 &       0.297   $\pm$   0.015   &         1304 $\pm$  6 &       2.121   $\pm$   0.022 \\ 
           867 $\pm$  8 &       0.300   $\pm$   0.012   &         1316 $\pm$  6 &       2.141   $\pm$   0.023 \\ 
           884 $\pm$  8 &       0.335   $\pm$   0.019   &         1328 $\pm$  6 &       2.203   $\pm$   0.027 \\ 
           900 $\pm$  8 &       0.362   $\pm$   0.011   &         1339 $\pm$  6 &       2.372   $\pm$   0.028 \\ 
           917 $\pm$  8 &       0.376   $\pm$   0.010   &         1350 $\pm$  6 &       2.379   $\pm$   0.030 \\ 
           933 $\pm$  8 &       0.413   $\pm$   0.010   &         1362 $\pm$  6 &       2.552   $\pm$   0.037 \\ 
           949 $\pm$  8 &       0.459   $\pm$   0.010   &         1374 $\pm$  6 &       2.574   $\pm$   0.026 \\ 
           965 $\pm$  8 &       0.498   $\pm$   0.010   &         1386 $\pm$  6 &       2.657   $\pm$   0.029 \\ 
           981 $\pm$  8 &       0.539   $\pm$   0.014   &         1399 $\pm$  6 &       2.765   $\pm$   0.027 \\ 
           997 $\pm$  8 &       0.607   $\pm$   0.010   &         1412 $\pm$  6 &       2.900   $\pm$   0.030 \\ 
          1013 $\pm$  8 &       0.672   $\pm$   0.010   &         1424 $\pm$  6 &       3.045   $\pm$   0.035 \\ 
          1028 $\pm$  8 &       0.763   $\pm$   0.011   &         1434 $\pm$  5 &       3.034   $\pm$   0.041 \\ 
\end{tabular}
\end{ruledtabular}
\end{table}
\end{document}